\documentclass[a4paper,11pt]{article}
\pdfoutput=1 

\usepackage{jheppub} 
\graphicspath{{./pics/}}

\usepackage[T1]{fontenc} 

\title{Critical points and number of master integrals}

\author{Roman N. Lee}
\author{and Andrei A. Pomeransky}
\affiliation{Budker Institute of Nuclear Physics and Novosibirsk State University, 630090 Novosibirsk, Russia}
\emailAdd{r.n.lee@inp.nsk.su}
\emailAdd{a.a.pomeransky@inp.nsk.su}

\abstract{
We consider the question about the number of master integrals for a multiloop Feynman diagram.
We show that, for a given set of denominators, this number is totally determined by the critical points of the polynomials entering
either of the two representations: the parametric representation and
the Baikov representation. In particular, for the parametric representation
the corresponding polynomial is just the sum of Symanzik polynomials.
The relevant topological invariant is the sum of the Milnor numbers
of the proper critical points. We present a \emph{Mathematica} package
\texttt{Mint} to automatize the counting of the master integrals.
}

\begin{document} 
\maketitle
\flushbottom

\section{Introduction}

\label{sec:intro}

Accurate theoretical predictions for the scattering amplitudes in
Standard Model and beyond require perturbative calculations at high
order. The feasibility of these calculations crucially depends on
our ability to evaluate the multiloop integrals. Remarkably, multiloop
integrals provide a fruitful ground for the application and development
of methods coming from various fields of mathematics, such as complex
analysis, differential and difference equations theory, algebraic
geometry etc. In a few last decades, an enormous progress in the
calculation of the multiloop integrals has been made. Deep insights
into the analytic and geometric nature of the multiloop integrals
have been gained. Many methods of the calculation have been invented.
However, each step up in the loop order is connected with a jump in
the computational complexity, so there is always a demand in new,
yet more powerful tools for the calculation of multiloop integrals.

One of the important tools that is relevant nowadays is the integration-by-part
(IBP) identities introduced in Refs. \cite{ChetTka1981,Tkachov1981}.
Using these identities, it is possible to reduce the calculation of
any multiloop integral with a given set of denominators to the calculation
of a finite number of the master integrals. An important feature of
this reduction is the possibility to construct differential and difference
equations for the master integrals. From the computational viewpoint,
the IBP reduction is known to be quite complicated problem. The reason
is the absence of the general effective algorithm for this reduction.
Almost all publicly available programs, like \texttt{FIRE} \cite{Smirnov2008}
and \texttt{Reduze} \cite{ManteuffelStuderus2012,Studeru2010}, heavily
rely on the Laporta algorithm, which includes a brute-force search
of the reduction rules. Recently one of the authors has presented
\texttt{LiteRed} package \cite{Lee2012} which performs a heuristic
search of the reduction rules. Nevertheless, up to now the problem
of the IBP reduction has not been solved.

This paper can be considered as a little step towards the construction
of the effective reduction algorithm. We present a simple recipe to
determine the number of the master integrals in the given sector.

In the next Section we show that it is possible to rewrite the parametric
representation in the form where the Symanzik polynomials $F$ and
$U$ enter only in the combination $F+U$. The integrals of the similar
form have been considered in Ref. \cite{AdolphsonSperber1997}. In
that paper the rank of the corresponding cohomology group has been
expressed via the volume of Newton polytope under some non-degeneracy
assumptions using the results of Ref. \cite{Kouchnirenko}. In Ref.
\cite{Kouchnirenko} the volume of Newton polytope has been related
to a certain topological invariant of a critical point, called Milnor
number (see the definition below). In Refs. \cite{fedoryuk1977method,Pham1981,Pham1985}
the homology group, connected with the Laplace integral, has been
considered. In these papers the independent cycles were related to
the steepest descent contours of the critical points of the Laplace
integral exponent. We combine the ideas of these papers to devise a simple algorithm
for counting the master integrals. Loosely speaking, it turns out
that this number is equal to the number of critical points of the
sum of Symanzik polynomials $F+U$. We demonstrate the efficiency
of this recipe on the example of a family of $4$-loop $g-2$ integrals.

\section{Parametric and Baikov representation}

Suppose that we are interested in the calculation of the $L$-loop
integral with $M$ denominators in $d$ dimensions depending on $E$
external momenta 
\begin{align}
J\left(\mathbf{n}\right) & =J(n_{1},n_{2},\ldots,n_{M})=\int\prod_{i=1}^{L}\frac{d^{d}l_{i}}{\pi^{d/2}}\prod_{\alpha=1}^{M}D_{\alpha}^{-n_{\alpha}}\,,\nonumber \\
D_{\alpha} & =A_{\alpha}^{ij}l_{i}\cdot l_{j}+2B_{\alpha}^{ik}l_{i}\cdot p_{k}+C_{\alpha}\,.
\end{align}
Here $p_{1},\ldots,p_{E}$ are linearly independent external momenta,
$A_{\alpha}$ are $L\times L$ matrices, $B_{\alpha}$ are $L\times E$
matrices, and $C_{\alpha}$ are some constants. 
\subsection*{Parametric representation}
The parametric representation
of $J\left(\mathbf{n}\right)$ has the form
\begin{equation}
J(\mathbf{n})=\frac{\Gamma\left(|\mathbf{n}|-Ld/2\right)}{\prod_{\alpha}\Gamma\left(n_{\alpha}\right)}\int\prod_{\alpha}dz_{\alpha}z_{\alpha}^{n_{\alpha}-1}\delta\left(1-\sum z\right)\frac{F^{Ld/2-|\mathbf{n}|}}{U^{\left(L+1\right)d/2-|\mathbf{n}|}}\,,\label{eq:FP}
\end{equation}
where $|\mathbf{n}|=\sum_{\alpha=0}^{M}n_{\alpha}$, $U$ and $F$
are the homogeneous polynomials of degrees $L$ and $L+1$, respectively.
These polynomials can be expressed in terms of quantities 
\begin{equation}
A^{ij}=\sum_{\alpha}z_{\alpha}A_{\alpha}^{ij},\quad B^{i}=\sum_{\alpha}z_{\alpha}B_{\alpha}^{ij}p_{j},\quad C=\sum_{\alpha}z_{\alpha}C_{\alpha}
\end{equation}
as follows 
\begin{equation}
U=\det\left(A\right),\quad F=\det\left(A\right)\, C-\left(A^{\mathrm{Adj}}\right)^{ij}B^{i}\cdot B^{j},
\end{equation}
where $A^{\mathrm{Adj}}=\det\left(A\right)A^{-1}$ is the adjoint
matrix.

Remarkably, it is possible to rewrite (\ref{eq:FP}) in the form,
which contains $U$ and $F$ only in the combination $F+U$. Indeed,
it is easy to show that the following representation holds: 
\begin{align}
J(\mathbf{n}) & =\frac{\Gamma\left(d/2\right)}{\Gamma\left(\left(L+1\right)d/2-|\mathbf{n}|\right)\prod_{\alpha}\Gamma\left(n_{\alpha}\right)}\int\limits _{0}^{\infty}\ldots\int\limits _{0}^{\infty}\prod_{\alpha}dz_{\alpha}z_{\alpha}^{n_{\alpha}-1}G^{-d/2}\,,\label{eq:FPm}\\
G & =F+U\,.
\end{align}
In order to pass from (\ref{eq:FPm}) to (\ref{eq:FP}) it is sufficient
to insert $1=\int ds\delta(s-\sum z)$, scale $z\to sz$ and integrate
over $s$.

Let us consider the integration-by-part identities in the representation
(\ref{eq:FPm}). The explicit form of these relations is not important
for the present consideration. We only note that, in contrast to the
momentum representation, the integration domain in Eq. (\ref{eq:FPm})
has a boundary (where some variables are equal to zero). Thus the
integration of a total derivative gives, in general, some surface
terms. These terms are expressed via the integrals in simpler sectors
(the integrals with some denominators missing). Therefore, if we want
to determine the master integrals in the given sector, we can safely
neglect these surface terms. 

\subsection*{Baikov representation}

So far we considered parametric representation (\ref{eq:FPm}) of
the multiloop integrals. Another representation for the multiloop
integrals has been introduced in Ref. \cite{Baikov1997}. Let us fix
the notation 
\begin{eqnarray*}
s_{ij} & = & q_{i}\cdot q_{j}\,,\\
q_{i} & = & \begin{cases}
l_{i}, & i\leqslant L\\
p_{i-L} & i>L
\end{cases}
\end{eqnarray*}
Then, the simplest way to derive the Baikov representation is to pass from
the integration over the loop momenta to the integration over 
\begin{equation}
s_{ij},\quad1\leqslant i\leqslant L,\quad i\leqslant j\leqslant L+E\,,\label{eq:svar}
\end{equation}
as described in Ref. \cite{Lee2010}. The total number of new variables
is $N=L\left(L+1\right)/2+LE$. Assuming that the denominators $D_{1},\ldots,D_{M}$
in Eq. (\ref{eq:FPm}) are linearly independent, we can choose $N-M$
irreducible numerators $D_{M+1},\ldots D_{N}$. The resulting formula
reads
\begin{align*}
J\left(\mathbf{n}\right) & =\frac{\pi^{\left(L-N\right)/2}S_{E}^{(E+1-d)/2}}{\Gamma\left[\left(d-E-L+1\right)/2,\ldots,\left(d-E\right)/2\right]}\\
 & \times\int\left(\prod_{i=1}^{L}\prod_{j=i}^{L+E}ds_{ij}\right)S^{(d-E-L-1)/2}\prod_{\alpha=1}^{N}D_{\alpha}^{-n_{\alpha}},
\end{align*}
where now $\mathbf{n}=\left(n_{1},\ldots n_{N}\right)$ and $n_{k>M}<0$.
The quantities $S$ and $S_{E}$ have the form 
\begin{align*}
S & =\det\left\{ \left.s_{ij}\right|_{i,j=1\ldots L+E}\right\} ,\quad S_{E}=\det\left\{ \left.s_{ij}\right|_{i,j=L+1\ldots L+E}\right\} \,.
\end{align*}
The functions $D_{\alpha}$ are linear functions of the variables
(\ref{eq:svar}), so that $\prod_{i=1}^{L}\prod_{j=i}^{L+E}ds_{ij}\propto dD_{1}\ldots dD_{N}$.
Thus, we have 
\begin{align*}
J\left(\mathbf{n}\right) & \propto\int\left(\prod_{\alpha=1}^{N}D_{\alpha}^{-n_{\alpha}}dD_{\alpha}\right)P^{(d-E-L-1)/2},
\end{align*}
where $P\left(D_{1},\ldots D_{N}\right)$ is obtained from $S$ by
expressing $s_{ij}$ via $D_{1},\ldots D_{N}$. This representation
is very similar to (\ref{eq:FPm}), except that now the variables
$D_{1},\ldots,D_{M}$ are raised to the negative powers. Following
the Baikov`s original idea, we choose the contours of integration
over these variables as sufficiently small circles around the origin
of the complex plane. In fact, this choice of the contours corresponds
to the maximal unitary cut of the integral. After this prescription,
the integrals in the subsectors are all vanishing. Taking the integrals
over $D_{1},\ldots,D_{M}$ by residues, we are left with the integrals
of the form
\begin{equation}
\int\left(\prod_{\alpha=M+1}^{N}D_{\alpha}^{-n_{\alpha}}dD_{\alpha}\right)P_{0}^{(d-I)/2},
\label{eq:GramP}
\end{equation}
where $P_{0}\left(D_{M+1},\ldots D_{N}\right)=P\left(0,\ldots,0,D_{M+1},\ldots,D_{N}\right)$,
and $I$ is some integer number.

So, one can see that both the parametric and Baikov representations can be written in the form depending on a single polynomial, $G$ and $P_0$, respectively.

\section{Number of master integrals,  basis of $M$-cycles and critical points}


For definiteness, let us consider here the parametric representation.
The integration-by-part identities determine
equivalence in the space of $J\left(\mathbf{n}\right)$, and master
integrals represent the basis in the quotient space, which is known
to be finite dimensional \cite{SmirPet2010}. Naturally, the question
about the number of master integrals arises. Due to the well-known duality
between the homology and cohomology groups, the dimensionality of
this quotient space, i.e., the number of master integrals, is equal
to the number of independent ``contours'' of integration, generating
no surface terms (and providing the convergence of the integral).
The homology group of these cycles has been considered by Pham in
Refs. \cite{Pham1981,Pham1985} and is equivalent to the relative
homology $H_{M}\left(\mathbb{C}^{M}\backslash\mathcal{Z},\mathcal{B}\right)$.
Here $\mathbb{C}^{M}\backslash\mathcal{Z}$ is a $2M$-dimensional
real variety obtained from $\mathbb{C}^{M}$ by removing algebraic
variety $\mathcal{Z}=\left\{ z\in\mathbb{C}^{M},G\left(z\right)=0\right\} $
and $\mathcal{B}=\left\{ z\in\mathbb{C}^{M},\left|G\left(z\right)\right|\geqslant B\right\} $
($B>0$ is large enough) is a set of points in $\mathbb{C}^{M}$ where
$\left|G\left(z\right)\right|$ is large enough. The number of master
integrals is the rank of $H_{M}\left(\mathbb{C}^{M}\backslash\mathcal{Z},\mathcal{B}\right)$.
In this section we specify the correspondence between the basis
cycles of $H_{M}\left(\mathbb{C}^{M}\backslash\mathcal{Z},\mathcal{B}\right)$
and the critical points of the polynomial $G$. One of the consequences
of this correspondence is the equality of the number of master integrals
and the sum of Milnor numbers of the proper critical points (see definition
below).

The above consideration is also valid for the Baikov representation with the replacement $G\to P_0$. 
In Ref. \cite{Baikov2006325} a criterion of the existence of master integral in a given sector
has been formulated. In our notations, this criterion states, that if the polynomial $P_{0}$ has no proper critical points, there is no master integrals in the sector.  It also states that the number of master integrals is bounded from below by the number of the nondegenerate isolated proper critical points of the polynomial $P_{0}$.
As far as it concerns the Baikov representation, our counting recipe can be considered as a developement of Ref. \cite{Baikov2006325}.

\subsection*{One-dimensional case}

\begin{figure}
\includegraphics[width=0.5\columnwidth]{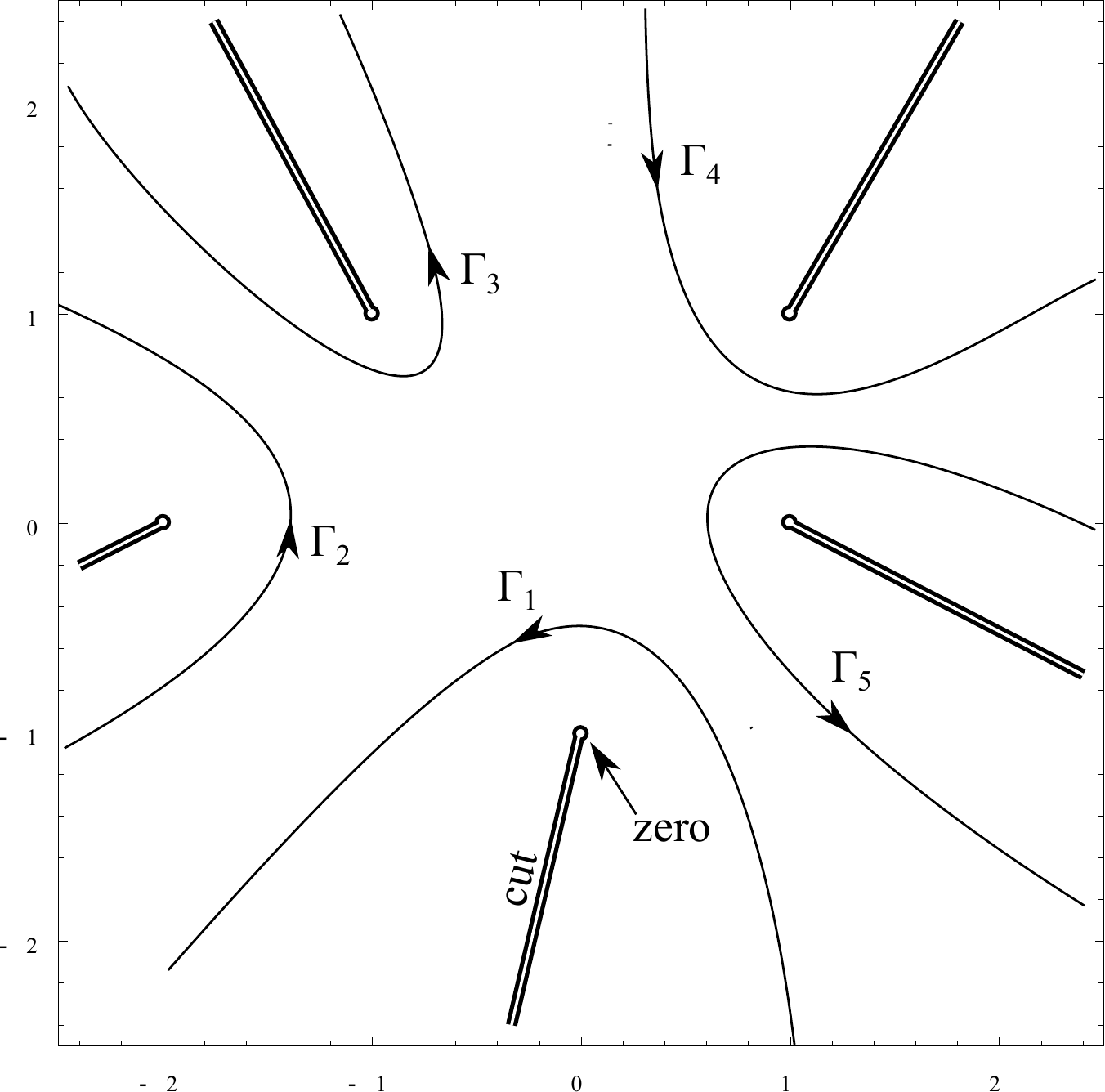}\protect\caption{Contour basis in the cut plane. Out of 5 contours $\Gamma_{1},\ldots\Gamma_{5}$
only 4 are independent, e.g. $\Gamma_{5}=-\Gamma_{1}-\Gamma_{2}-\Gamma_{3}-\Gamma_{4}$}
\label{fig:contours-1}
\end{figure}

We consider the integral 
\begin{equation}
\int\frac{dz\ z^{n-1}}{G\left(z\right)^{\nu}},\label{eq:int}
\end{equation}
where $G\left(z\right)$ is some polynomial of $p$-th degree of a
single variable $z$. The integrand is defined in the cut plane with
cuts starting at zeros of $G\left(z\right)$ and going to infinity.
We want to determine the number of independent contours of integration
which do not give rise to the surface terms. Obviously, for large
enough positive $\nu$ those contours should start and end at infinity,
embracing one or more cuts. Of course, the result is known in advance:
this number is one less than the number of distinct zeros of the polynomial
$G\left(z\right)$. This statement is demonstrated in Fig. \ref{fig:contours-1}.
However, we would like to describe an approach which can be generalized
to the case of many variables. 

Let $z_{0}^{\left(1\right)},\ldots,z_{0}^{\left(k\right)}$ are distinct
zeros of $G$ with degeneracies $p^{\left(1\right)},\ldots,p^{\left(k\right)}$,
so that $\sum_{i}p^{\left(i\right)}=p$. Then, obviously, $z_{0}^{\left(i\right)}$
is also zero of $\partial_{z}G$ with degeneracy $p^{\left(i\right)}-1$
if $p^{\left(i\right)}>1$. If $z_{0}^{\left(i\right)}$ is not degenerate
($p^{\left(i\right)}=1$), then necessarily $\partial_{z}G\left(z_{0}^{\left(i\right)}\right)\neq0$.
Then, out of $p-1$ zeros of $\partial_{z}G$ (called\emph{ critical
points} of $G$ in what follows) there are exactly $p-1-\sum_{i}\left(p^{\left(i\right)}-1\right)=k-1$
critical points which are not zeros of $G$. In what follows we will
call them \emph{proper critical points}. So, the number of independent
contours is equal to the number of proper critical points of $G$
(including degeneracy). This simple observation hints to a deep connection
between independent contours and proper critical points. 

To reveal this connection, let us assume that all proper critical
points $z^{\left(1\right)},\ldots z^{\left(k-1\right)}$ are non-degenerate
and all \emph{critical phases }$\phi^{\left(i\right)}=\arg G\left(z^{\left(i\right)}\right)$
are distinct. Let us first consider the curves in the complex plane
of $z$ defined by the condition $\arg G\left(z\right)=\phi$, where
$\phi$ is some noncritical phase. From the Cauchy-Riemann condition,
these curves are gradient flow curves of $h\left(z\right)=\ln\left|G\left(z\right)\right|$.
For given $\phi$ one can draw such a curve starting from each zero
and going to infinity. These curves provide a natural choice for the
cuts. Let us now consider the curves defined by the condition $\arg G\left(z\right)=\phi^{\left(i\right)}$,
where $\phi^{\left(i\right)}$ is the critical phase. The corresponding
critical point $z^{i}$ is a saddle point of $h\left(z\right)$, i.e.
it is an intersection of the curves of the steepest descent $\Gamma_{-}\left(z^{\left(i\right)}\right)$
and of the steepest ascent $\Gamma_{+}\left(z^{\left(i\right)}\right)$.
The curve $\Gamma_{-}\left(z^{\left(i\right)}\right)$ obviously ends
at zeros, while $\Gamma_{+}\left(z^{\left(i\right)}\right)$ is going
to infinity never passing through zero. Let us consider a superposition
$\Gamma=\sum c_{i}\Gamma_{+}\left(z^{\left(i\right)}\right)$. The
integer coefficient $c_{i}$ is equal to the intersection index of
$\Gamma$ with the contour $\Gamma_{-}\left(z^{\left(i\right)}\right)$
for a suitable choice of the orientation of $\Gamma_{\pm}\left(z^{\left(i\right)}\right)$.
The intersection index is topological invariant, i.e., it can not
be changed by continuous deformations (in the cut plane). Therefore,
$\Gamma\thicksim0$ (is contractible) only when all $c_{i}$ are zero,
which means that the contours 
\[
\Gamma_{1}=\Gamma_{+}\left(z^{\left(1\right)}\right),\ldots,\Gamma_{k-1}=\Gamma_{+}\left(z^{\left(k-1\right)}\right)
\]
are independent. In one-dimensional case the completeness of this
set is obvious and therefore this system forms a basis. Note that
for negative $\nu$ the basis of contours can be chosen as the set
of $\Gamma_{-}\left(z^{\left(i\right)}\right)$.

\begin{figure}
\includegraphics[width=0.5\columnwidth]{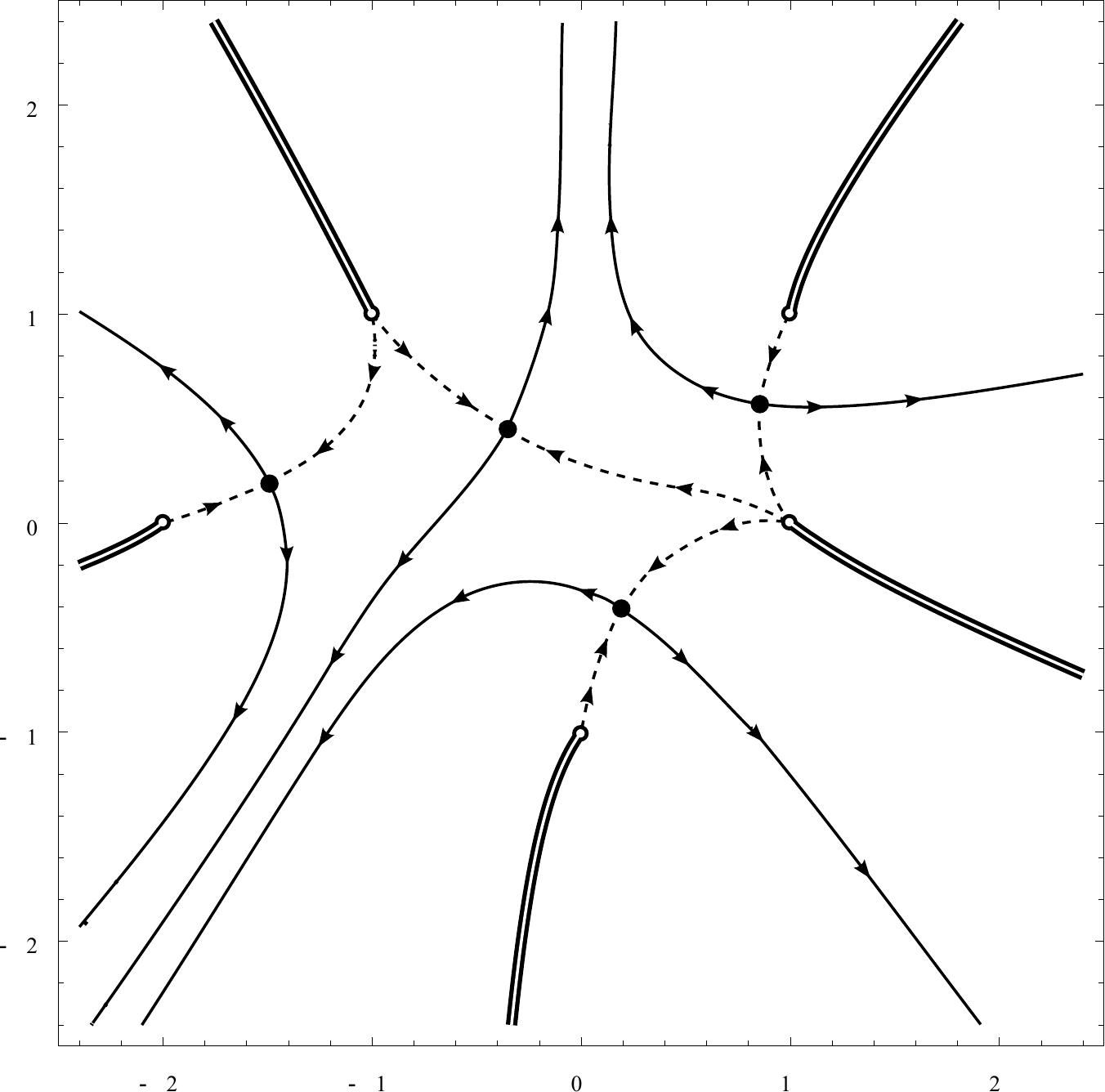}\protect\caption{Saddle-point contours $\Gamma_{+}\left(z^{(i)}\right)$ and $\Gamma_{-}\left(z^{(i)}\right)$ in the cut plane (respectively, solid and dashed curves with arrows).}
\label{fig:contours2-1}
\end{figure}

\subsection*{Multidimensional case}

Let us now briefly consider the multidimensional case. We have the
integral (\ref{eq:int}), where now $z=\left(z_{1},\ldots,z_{M}\right)$
and $dz z^{n-1}=dz_{1}z_1^{n_1-1}\ldots dz_{M}z_{M}^{n_1-1}$. We want to determine the number of
independent multidimensional ``contours'' of integration which are
$M$-cycles in $\mathbb{C}^{M}$ space.

Zeros of $G\left(z\right)$ are no more isolated points, but hypersurfaces
of $M-1$ complex dimensions. Remarkably, the solution of $M$ complex
equations for gradient $\partial G/\partial z_{\alpha}=0$ consists,
in non-degenerate case, of some isolated points, which we again call
critical points of $G\left(z\right)$. We consider non-degenerate
case in the following sense:
\begin{enumerate}
\item There is a finite number of proper critical points $z^{\left(1\right)},\ldots,z^{\left(k-1\right)}$
defined as the solutions of 
\begin{align}
\partial G/\partial z_{\alpha} & =0,\quad\alpha=1,\ldots,N\nonumber \\
G\left(z\right) & \neq0\label{eq:proper_critical_points}
\end{align}

\item The Hessian matrix $\frac{\partial^{2}G}{\partial z_{\alpha}\partial z_{\beta}}\left(z^{\left(i\right)}\right)$
at each critical point is invertible.
\item The critical phases $\phi^{\left(i\right)}=\arg G\left(z^{\left(i\right)}\right)$
are all distinct.
\end{enumerate}
Now the consideration of the previous subsection can be easily generalized.
We can follow a usual construction of Morse theory using $h\left(z\right)=\ln\left|G\left(z\right)\right|$
as a Morse function and $x_{\alpha}=\Re z_{\alpha},\, y_{\alpha}=\Im z_{\alpha}$
as coordinates. The Morse theory is formulated for a Riemannian manifold,
which in our case has a flat metrics. Then the gradient flow equations
have the form 
\begin{eqnarray}
\frac{dx_{\alpha}}{dt} & = & \frac{\partial h}{\partial x_{\alpha}}\,,\label{eq:grad-flow}\\
\frac{dy_{\alpha}}{dt} & = & \frac{\partial h}{\partial y_{\alpha}}\,.\nonumber 
\end{eqnarray}
This system can be written in the form
\begin{equation}
\frac{dz_{\alpha}}{dt}=\frac{\partial h}{\partial\bar{z}_{\alpha}}\,.\label{eq:grad-flowc}
\end{equation}
We determine $\Gamma_{\pm}\left(z^{\left(i\right)}\right)$ as the
union of the trajectories of Eqs. (\ref{eq:grad-flowc}) subject to
the condition $z\left(t\right)\stackrel{t\to\mp\infty}{\longrightarrow}z^{\left(i\right)}$.
The varieties $\Gamma_{\pm}\left(z^{\left(i\right)}\right)$ are nothing
but the Lefschetz thimbles \cite{lefschetz663analysis,Pham1985},
see also Ref. \cite{Witten2010}.  Due to Cauchy-Riemann conditions,
the phase of $G\left(z\right)$ on $\Gamma_{\pm}\left(z^{\left(i\right)}\right)$
remains constant and is equal to $\phi_{i}$, so that the contours
$\Gamma_{+}\left(z^{\left(i\right)}\right)$ and $\Gamma_{-}\left(z^{\left(j\right)}\right)$
intersect only for $i=j$. Then the independence of $\Gamma_{+}\left(z^{\left(i\right)}\right)$
can be proved in the same way as in 1d case. In fact, it is known,
that Lefschetz thimbles constitute the basis of the relative homology
$H_{M}\left(\mathbb{C}^{M}\backslash\mathcal{Z},\mathcal{B}\right)$,
see Ref. \cite{Pham1985}. In particular, it means that the rank of
the relative homology group is equal to the number of critical points
if conditions 1-3 are satisfied. 

If the conditions 2,3 are not fulfilled for $G$, we can perform a
small perturbations $G\left(z\right)\to G_{\epsilon}\left(z\right)=G\left(z\right)+\epsilon g\left(z\right)$,
where $g\left(z\right)$ is some suitable polynomial and consider
only those critical points of $G_{\epsilon}\left(z\right)$, which
are close to $z^{\left(i\right)}$. The number of the critical points
of $G_{\epsilon}\left(z\right)$ in the vicinity of $z^{\left(i\right)}$
is the ``multiplicity'' of $z^{\left(i\right)}$, an invariant called
Milnor number of $G\left(z\right)$ at $z=z^{\left(i\right)}$, see,
e.g. Ref. \cite{AGV}. So, we come to the following conclusion: \emph{If
$G\left(z\right)$ has only isolated proper critical points, the number
of independent contours of integration ($M$-cycles) is equal to the
sum of Milnor numbers of the proper critical points of $G\left(z\right)$.} 

Zeros of the polynomial $G$ are the branching points of the integrand
in Eq. (\ref{eq:FPm}). The cuts can be chosen as $\left(2M-1\right)$
real dimensional variety $C_{\phi}$ determined by the condition $\arg G\left(z\right)=\phi$,
where $\phi$ is a fixed noncritical value of the phase. Note that
the contours $\Gamma_{+}\left(z^{\left(i\right)}\right)$ do not intersect
the cuts.

Let us consider now the case when $G\left(z\right)$ has non-isolated
proper critical points. It means that the set of points, where Eq.
(\ref{eq:proper_critical_points}) is satisfied, forms a critical variety
of dimension $\geq1$. In practical applications, as illustrated in
Sec. \ref{sec:Example:-4-loop-onshell}, non-isolated proper critical
points are rather rare. In this case one can still construct the basis
of $M$-cycles, but it requires somewhat more work. The solution of
Eq. (\ref{eq:proper_critical_points}) is a union of several algebraic
varieties --- the irreducible parts. Some of these irreducible parts
may be isolated points, and they should be treated as explained above. 

Let $\mathcal{V}$ be an irreducible component of dimension $s>0.$
In order to construct the $M$-cycles passing through $\mathcal{V}$
one has to consider the compact $s$-cycles of $\mathcal{V}$ (the
elements of the middle homology group). For each $s$-cycle from the
basis of this homology group one has to consider the union of upward
gradient flow lines $\Gamma_{+}$ starting on the points of that cycle.
For each point of the cycle, these lines form a variety of dimension
$M-s$. Altogether, these lines form the $M$-cycle, a member of the
basis we are looking for. This consideration assumes non-degeneracy
of $\mathcal{V}$. However, it can happen, that the critical variety
$\mathcal{V}$ is degenerate, that is the Hessian matrix has zero
modes which are not tangent to $\mathcal{V}$. In this case there
are several $M$-cycles per each $s$-cycle. To sum up, the number
of the independent $M$-cycles is equal to the sum of Milnor numbers
of isolated proper critical points plus the number of independent
$s$-cycles, on the $s$-dimensional components ($s>0$) of the critical
set (counted with multiplicity, if the component is degenerate). Let
us iterate, that in the applications to multiloop integrals the non-isolated
critical points appear very rarely, and the more so do the degenerate
non-isolated critical points.

In conclusion of this Section we note that similar ideas appeared
earlier in Refs. \cite{Marcolli2008,marcolli2010feynman}. The difference
with our approach is that in Refs. \cite{Marcolli2008,marcolli2010feynman}
the critical points of Symanzik polynomials $U$ and $F$ were studied
separately. These points tend to be non-isolated, which makes their
treatment more difficult. In order to apply this approach to counting
the master integrals, it is necessary to consider the critical points
of the map $(U$, $F)$ : $\mathbb{C}^{M}\rightarrow\mathbb{C}^{2}$,
with the additional condition $U=1$. It can be shown that this approach,
up to some details, is equivalent to one presented above.

\subsection*{Pedagogical example}

From the above consideration it follows that we can count the number
of master integrals in a given sector by counting the number of proper
critical points (accounting for their possible multiplicity) of the
polynomial $F+U$. As an example, let us consider the following family
of sunrise integrals
\[
J\left(n_{1},n_{2},n_{3}\right)=\int\frac{\mathrm{d}^{d}l_{1}\mathrm{d}^{d}l_{2}}{\left(i\pi^{d/2}\right)^{2}}\left[l_{1}^{2}+1\right]^{-n_{1}}\left[l_{2}^{2}+1\right]^{-n_{2}}\left[\left(l_{1}+l_{2}-p\right)^{2}+1\right]^{-n_{3}}
\]

where $n_{i}\in\mathbb{N}$. Note that we need not introduce irreducible
numerators. We have
\begin{align*}
J(n_{1},n_{2},n_{3}) & =\frac{\Gamma\left(d/2\right)\iiint\limits _{0}^{\infty}dz_{1}z_{1}^{n_{1}-1}dz_{2}z_{2}^{n_{2}-1}dz_{3}z_{3}^{n_{3}-1}G^{-d/2}}{\Gamma\left(3d/2-n_{1}-n_{2}-n_{3},n_{1},n_{2},n_{3}\right)}\,,\\
G & =F+U=z_{1}z_{2}+z_{1}z_{3}+z_{2}z_{3}\\
 & +z_{1}^{2}z_{3}+z_{2}^{2}z_{3}+z_{1}z_{2}^{2}+z_{1}z_{3}^{2}+z_{2}z_{3}^{2}+z_{1}^{2}z_{2}+\left(p^{2}+3\right)z_{1}z_{2}z_{3}\,.
\end{align*}

\begin{figure}
\includegraphics[width=1\textwidth]{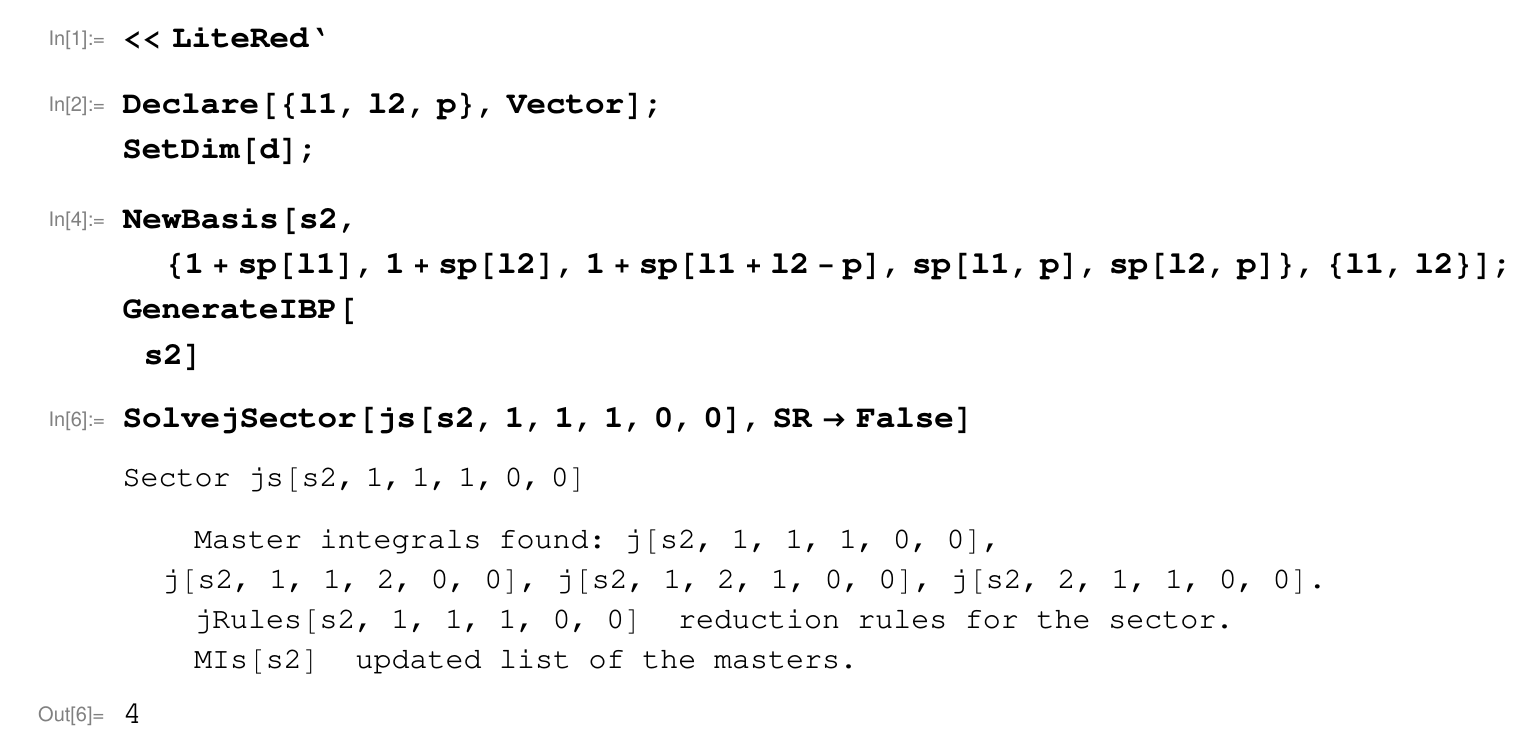}\protect\caption{Finding reduction rules for $J\left(n_{1},n_{2},n_{3}\right)$ with
\texttt{LiteRed}.}
\label{fig:LiteRedReduction}
\end{figure}

The polynomial $G$ has eight critical points (the solutions of $\nabla G=0$)
\begin{gather*}
z^{\left(1\right)}=-\frac{\left(p^{2}-1,1,1\right)}{3\left(p^{2}+1\right)},\ z^{\left(2\right)}=-\frac{\left(1,p^{2}-1,1\right)}{3\left(p^{2}+1\right)},\ z^{\left(3\right)}=-\frac{\left(1,1,p^{2}-1\right)}{3\left(p^{2}+1\right)},z^{\left(4\right)}=-\frac{2\left(1,1,1\right)}{p^{2}+9},\\
z^{\left(5\right)}=(0,0,-1),\quad z^{\left(6\right)}=(0,-1,0),\quad z^{\left(7\right)}=(-1,0,0),\quad z^{\left(8\right)}=(0,0,0)
\end{gather*}
of which the first four are nonzero. All four nonzero critical points
$z^{\left(1\right)},\ldots,z^{\left(4\right)}$ are non-degenerate,
therefore there are four master integrals in this sector. Indeed,
running the simple \emph{Mathematica} program using\texttt{ LiteRed}
package, see Fig. \ref{fig:LiteRedReduction} reveals four master
integrals. 

Note the option \textbf{$\mathrm{\mathbf{SR}\to\mathbf{False}}$},
which forbids \texttt{LiteRed} to use symmetries of the integral with
respect to permutations of indices. If we used the default setting
$\mathrm{\mathbf{SR}\to\mathbf{True}}$ instead, \texttt{LiteRed}
would clearly find only two independent integrals $J\left(1,1,1\right)$
and $J\left(2,1,1\right)$. It is easy to account for the permutation
symmetries also for critical points $z^{\left(1\right)},\ldots,z^{\left(4\right)}$. Namely, there are two orbits of the permutation group acting on
the critical points: $\{z^{\left(4\right)}\}$ and $\{z^{\left(1\right)},z^{\left(2\right)},z^{\left(3\right)}\}$.
We see that the number of orbits gives the number of master integrals
with the account of symmetries.

In order to use the Baikov representation, we need to introduce two irreducible
numerators $D_{4},D_{5}$. This can be done quite arbitrarily, provided
that, together with $D_{1}=l_{1}^{2}+1,\, D_{2}=l_{2}^{2}+1,\, D_{3}=\left(l_{1}+l_{2}-p\right)^{2}+1$,
they form a complete basis. We choose
\[
D_{4}=\left(l_{1}-p\right)^{2},\quad D_{5}=\left(l_{2}-p\right)^{2}
\]

We have 
\[
P_{0}=\frac{D_{4}D_{5}}{4}\left(p^{2}-3-D_{4}-D_{5}\right)+\frac{\left(p^{2}+1\right)^{2}}{4}
\]
 There are four critical points $z^{\left(i\right)}=\left(D_{4}^{\left(i\right)},D_{5}^{\left(i\right)}\right)$:
\begin{gather*}
z^{\left(1\right)}=\left(0,0\right),\ z^{\left(2\right)}=\left(0,p^{2}-3\right),\ z^{\left(3\right)}=\left(p^{2}-3,0\right),z^{\left(4\right)}=\frac{1}{3}\left(p^{2}-3,p^{2}-3\right)\,.
\end{gather*}
 Each point is non-degenerate and proper, so we again conclude that
there are 4 master integrals if we neglect the symmetry relations. 

The account of the symmetry relations is somewhat less obvious in
this representation than in the parametric representation. The reason
is that the numerators are transformed one-to-many upon the symmetry,
in contrast to the one-to-one transformation of the denominators.
Nevertheless, we can find the action of the symmetries on $D_{4}$
and $D_{5}$ from the corresponding mapping of loop momenta. In the
linear subspace determined by $D_{1}=D_{2}=D_{3}=0$ they read 
\begin{align*}
1. & D_{4}\to D_{4},\ D_{5}\to D_{5}\\
2. & D_{4}\to D_{5},\ D_{5}\to D_{4}\\
3. & D_{4}\to D_{4},\ D_{5}\to p^{2}-3-D_{4}-D_{5}\\
4. & D_{4}\to D_{5},\ D_{5}\to p^{2}-3-D_{4}-D_{5}\\
5. & D_{4}\to p^{2}-3-D_{4}-D_{5},\ D_{5}\to D_{5}\\
6. & D_{4}\to p^{2}-3-D_{4}-D_{5},\ D_{5}\to D_{4}
\end{align*}

Now it is trivial to find the orbits of this symmetry group: they
are $\{z^{\left(4\right)}\}$ and $\{z^{\left(1\right)},z^{\left(2\right)},z^{\left(3\right)}\}$.
Therefore, we again conclude, that after the account of the symmetries,
there are two master integrals.

\section{Algebraic treatment}

There is a well-known correspondence between algebraic varieties and
ideals in the polynomial rings. Due to this correspondence, in order to find
the sum of Milnor numbers of proper critical points, one need not
explicitly solve the polynomial system of equations \ref{eq:proper_critical_points}.
Instead, one may calculate the dimensionality of the quotient ring
of the polynomial ideal we will describe in a moment. If we were interested
in the sum of Milnor numbers of all critical points (including non-proper
ones), we would choose Jacobian ideal, which is generated by $\partial G/\partial z_{1},\ldots,\partial G/\partial z_{M}$.
The condition $G\neq0$ can be taken into account by introducing an
extra variable $z_{0}$ and considering the ideal 
\begin{equation}
\mathcal{I}=\langle\partial G/\partial z_{1},\ldots,\partial G/\partial z_{M},z_{0}G-1\rangle\,.\label{eq:ideal}
\end{equation}
Choosing some monomial ordering and constructing the Groebner basis,
we can determine the set of irreducible monomials. Then the dimensionality
of the quotient space is just the number of those monomials.

\subsection*{Symmetries}

The above method gives us the number of master integrals without the
account of possible symmetry relations between them. In terms of the
polynomial $G$, those symmetries are such permutations of the variables
$z_{1},\ldots z_{M}$, which leave $G$ intact. They form a permutation
group $P$. 

The most straightforward way to take those symmetries into account
is the following. For each irreducible monomial $m=z_{1}^{n_{1}}z_{2}^{n_{2}}\ldots z_{M}^{n_{M}}$
and for each permutation $p=\left(p_{1},p_{2},\ldots,p_{M}\right)\in P$
we construct and reduce with respect to $\mathcal{I}$ the polynomial
$m-pm=z_{1}^{n_{1}}z_{2}^{n_{2}}\ldots z_{M}^{n_{M}}-z_{p_{1}}^{n_{1}}z_{p_{2}}^{n_{2}}\ldots z_{p_{M}}^{n_{M}}$.
Let us denote as $r$ the number of linearly independent remainders. Then, the number of the
master integrals surviving the symmetry relations is just the number
of irreducible monomials minus $r$.

\subsection*{Non-isolated critical points.}

The existence of non-isolated proper critical points can be easily
seen as infinite dimensionality of the quotient ring of $\mathcal{I}$.
Then one needs to determine the irreducible components of the critical
set, together with their possible degeneracy. This problem can be
naturally solved with algebraic approach by performing the primary
decomposition of $\mathcal{I}$, i.e., finding the decomposition
\[
\mathcal{I}=\bigcap\mathcal{I}_{i}\,,
\]
where each $\mathcal{I}_{i}$ is a primary ideal corresponding to
some irreducible component. If $\mathcal{I}_{i}$ is also prime, the
corresponding critical variety is non-degenerate. The number of middle-dimensional
cycles in this critical variety can be also determined by the algebraic
method (see next Section), but the discussion of the most general
case is beyond the scope of this paper.

\subsection*{\textit{Mathematica} package\texttt{ Mint}}

\begin{figure}
\includegraphics[width=1\textwidth]{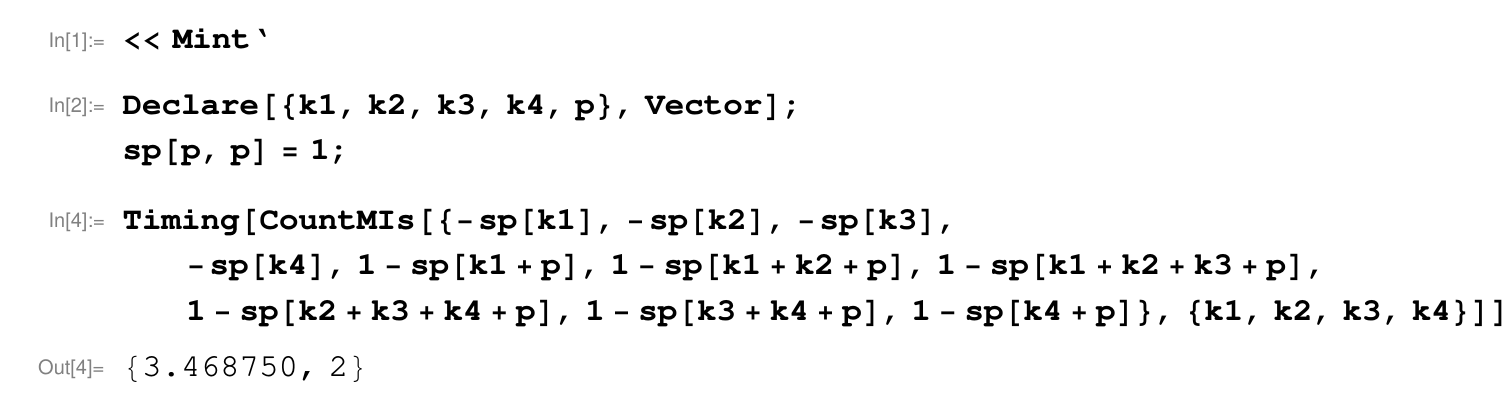}\protect\caption{Example of using\texttt{ Mint}.}
\label{fig:MintDemo1}
\end{figure}

\begin{figure}
\includegraphics[width=1\textwidth]{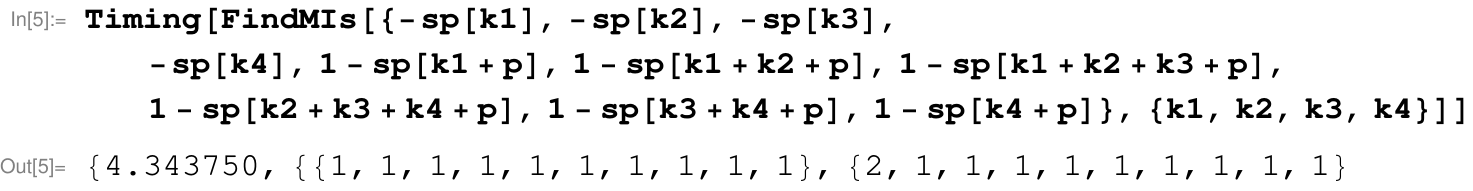}\protect\caption{Finding the master integrals with \textbf{$\mathbf{FindMIs}$}.}
\label{fig:MintDemo2}
\end{figure}

\begin{figure}
\includegraphics[width=1\textwidth]{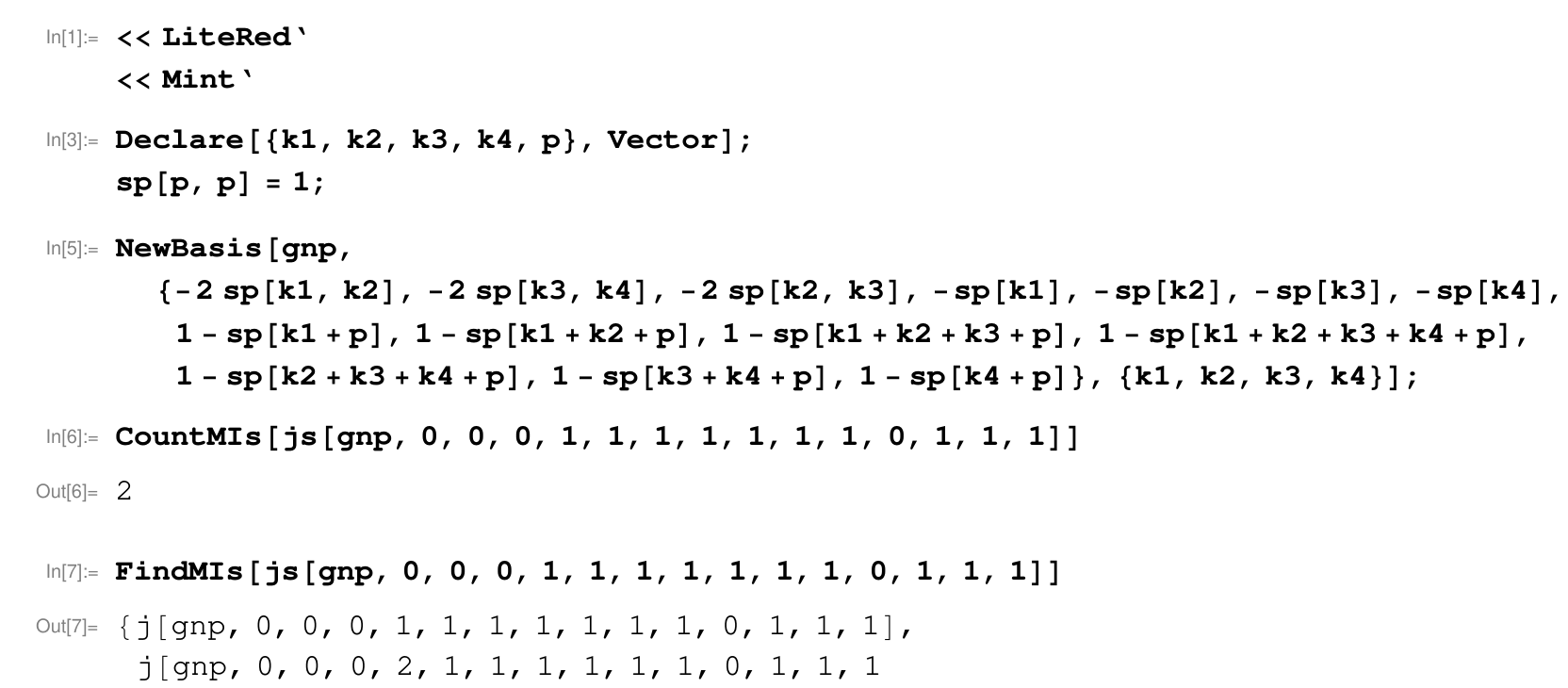}\protect\caption{Example of using\texttt{ Mint} together with \texttt{LiteRed}.}
\label{fig:MintDemo3}
\end{figure}

From the above consideration it follows, that, apart from the case
of non-isolated critical points, the problem of determination of the
number of independent $M$-cycles (equal to the number of master integrals)
can be easily solved on any modern computer algebra system. We have
developed a simple \textit{Mathematica} package \texttt{Mint}, \cite{Mint}, which
finds the number of the master integrals with a given set of denominators
if the proper critical points of the corresponding polynomial $F+U$
are isolated. The example of its usage is shown in Fig.\ref{fig:MintDemo1}
 Note that, knowing the number $M$ of master integrals (with account
of symmetry) in a given sector, one can select as masters almost any
$M$ integrals, provided that no two of them are equal due to symmetry
relation. The accidental linear dependency between them, though may
happen in principle, is very unlikely. The \texttt{Mint} package contains
a procedure \textbf{$\mathbf{FindMIs}$ }which suggests the simplest
integrals which can be chosen as masters. Its output is a list of
multi-indices, corresponding to a possible choice of the master integrals,
see Fig. \ref{fig:MintDemo2}. 
 If \texttt{Mint} is loaded after \texttt{LiteRed}, the procedures
\textbf{$\mathbf{CountMIs}$ }and \textbf{$\mathbf{FindMIs}$ }can
be called directly for the sectors, see the example in Fig. \ref{fig:MintDemo3}

By default, the \texttt{Mint} package uses parametric representation
to count the master integrals. If used with \texttt{LiteRed}, it can
also rely on the Baikov representation. Presumably, this approach should
be useful for higher sectors, when the number of numerators is small.
The corresponding call of the procedures \textbf{$\mathbf{CountMIs}$
}and \textbf{$\mathbf{FindMIs}$ }should include option $\mathbf{Method}\to"\mathbf{GramP}"$.
Before this call, the \texttt{LiteRed}`s procedure $\mathbf{FindSymmetries}$
should be called in order to determine the symmetries of the numerators.
We should notice that both methods, the one based on parametric representation
and the one based on the Baikov representation worked equally effective for
the complicated cases, such as the one described in the next Section.
Moreover, the non-isolated critical points seem to appear simultaneously
in both approaches.

\section{Example: 4-loop onshell $g-2$ integrals\label{sec:Example:-4-loop-onshell}}

\begin{figure}
\includegraphics[width=0.6\textwidth]{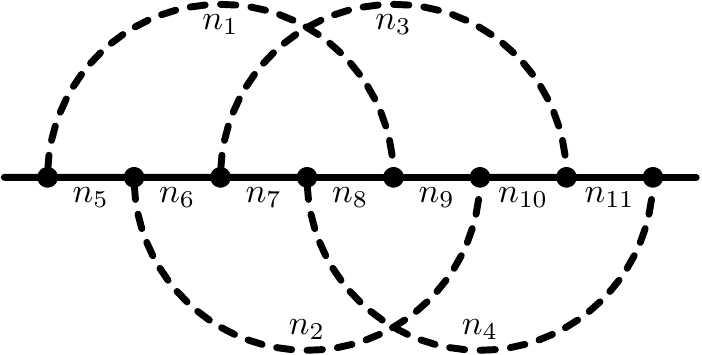}\protect\caption{The family of integrals considered.}
\label{fig:gnp}
\end{figure}

As a nontrivial example of the application of the above method, let
us consider the family of the integrals shown in Fig. \ref{fig:gnp}.

\begin{figure}
\includegraphics[width=1\textwidth]{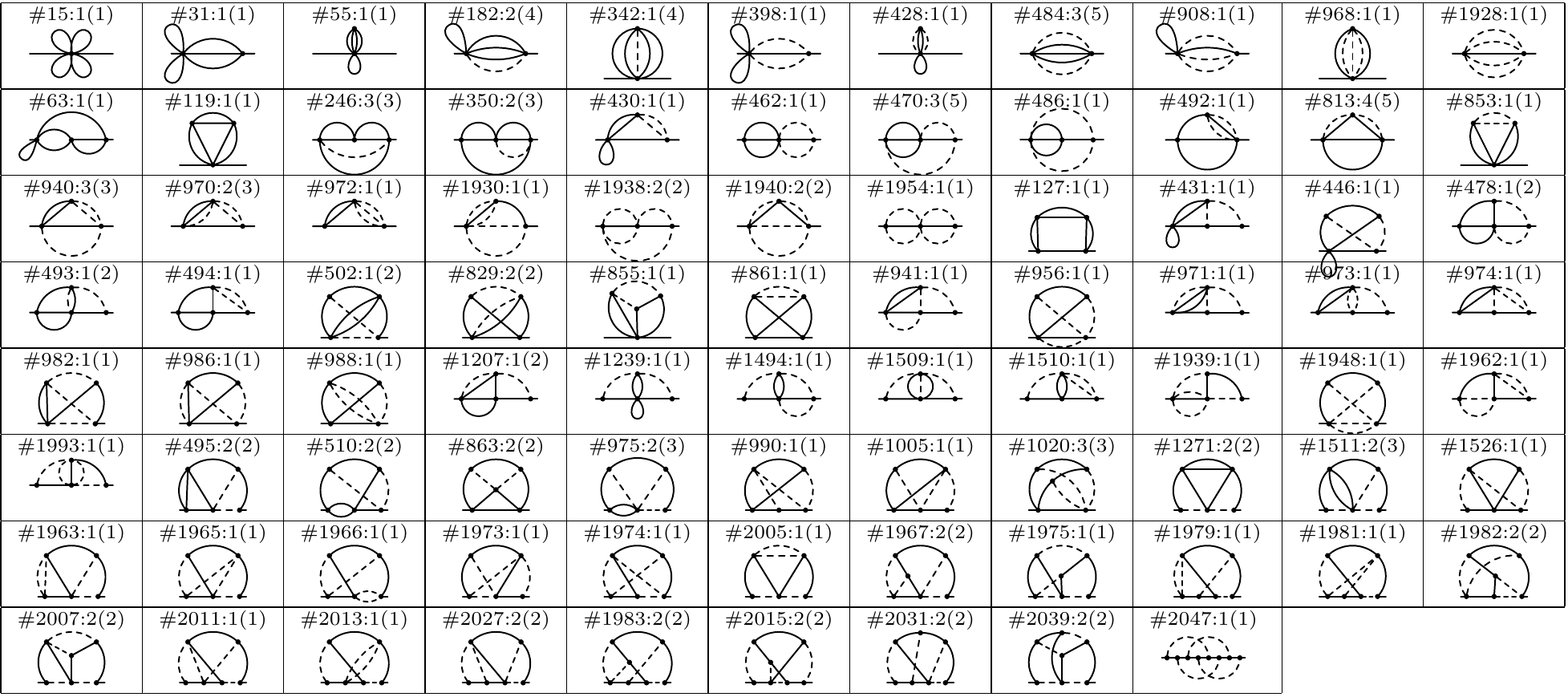}\protect\caption{Master integrals. }
\label{fig:mis}
\end{figure}

There are $261$ nonzero nonequivalent sectors in this family. Application
of our counting method gives $84$ sectors containing in total $119$
master integrals. The graphs for each sector, together with the number
of master integrals, are shown in Fig. \ref{fig:mis}. The sectors
are numbered in the following way: for each sector its number is the
string of indices of its simplest integral understood as binary number.
E.g., $\#350=00101011110_{2}$ corresponds to the integrals with denominators
$d_{3},d_{5},d_{7},d_{8},d_{9},d_{10}$. The label above each diagram
should be read as follows: $\#350:2(3)$ means that the sector $\#350$
has $2$ masters ($3$ masters) if the symmetries are used (not used).

\subsection*{Treating non-isolated critical points}

Almost for all nonzero nonequivalent sectors the polynomials $G=F+U$
have isolated proper critical points. Out of 261 sectors there are
7 exceptions: sectors $\#246$, $\#350$, $\#414$, $\#429$, $\#821$,
$\#924$, and $\#969$. In each case there is a 1-dimensional critical
variety. Let us explain how we determined the number of independent
$M$-cycles for these cases on the example of sector $\#350$. For
this sector we have
\begin{align*}
G & =F+U=\left(z_{2356}+1\right)\left(y_{1235}+y_{1236}+y_{1256}+y_{1356}+y_{2356}\right)\\
 & +z_{4}z_{25}\left(z_{23456}+1\right)\left(y_{13}+y_{16}+y_{36}\right)\,,
\end{align*}
where we have used the abbreviations $z_{ij\ldots k}=z_{i}+z_{j}+\ldots z_{k}$,
$y_{ij\ldots k}=z_{i}z_{j}\ldots z_{k}$. The quotient algebra for
the ideal $\mathcal{I}$, Eq. (\ref{eq:ideal}), is infinite dimensional,
which indicates non-isolated critical points. Then we calculate the
primary decomposition of $\mathcal{I}$, e.g., by using \texttt{Sage}
\cite{sage}. We have
\[
\mathcal{I}=\mathcal{I}_{1}\cap\mathcal{I}_{2},
\]
where
\begin{align*}
\mathcal{I}_{1} & =\left\langle 5z_{5}+1,\,5z_{4}+1,\,5z_{3}+5z_{6}+1,\,5z_{2}+1,\,5z_{1}-1,\,4z_{0}+3125,\,25z_{6}^{2}+5z_{6}-1\right\rangle \,,\\
\mathcal{I}_{2} & =\left\langle 10z_{6}+3,\,10z_{3}+3,\,5z_{2}+5z_{5}+1,\,20z_{1}-3,27z_{0}+50000,\right.\\
 & \left.100z_{4}^{2}+100z_{5}^{2}+20z_{4}+20z_{5}-3\right\rangle \,.
\end{align*}
Both ideals are prime. The quotient space of the first ideal is 2-dimensional,
in accordance with the fact, that the corresponding polynomial system
has two solutions:

\begin{align*}
z^{\left(1\right)} & =\frac{1}{5}\left(1,-1,1/\varphi,-1,-1,-\varphi\right)\,,\\
z^{\left(2\right)} & =\frac{1}{5}\left(1,-1,-\varphi,-1,-1,1/\varphi\right)\,,
\end{align*}
where $\varphi=\left(\sqrt{5}+1\right)/2$. The quotient space of
$\mathcal{I}_{2}$ is infinite dimensional. The middle homology of
the algebraic variety determined by $\mathcal{I}_{2}$ obviously coincides
with that of the variety in $\mathbb{C}^{2}$ determined by the equation
$\tilde{G}\left(z_{4},z_{5}\right)=0$, where 
\[
\tilde{G}\left(z_{4},z_{5}\right)=100z_{4}^{2}+100z_{5}^{2}+20z_{4}+20z_{5}-3\,.
\]
Remarkably, the basis of this homology can be found by exactly the
same method that we used before, see, e.g. \cite{AGV}. In fact, the
homology basis is formed by the cycles (called the vanishing cycles)
which are the intersection of the Lefschetz thimbles with the variety
determined by $\tilde{G}\left(z_{4},z_{5}\right)=0$. We simply calculate
the dimensionality of the quotient space of the ideal
\[
\tilde{\mathcal{I}}=\langle\partial\tilde{G}/\partial z_{4},\partial\tilde{G}/\partial z_{5},z_{0}\tilde{G}-1\rangle\ .
\]
This dimensionality is equal to 1, which corresponds to one solution
\[
\tilde{z}^{\left(3\right)}=-\frac{1}{10}\left(1,1\right)\,.
\]
In total we have two $M$-cycles passing through $z^{\left(1\right)}$
and $z^{\left(2\right)}$ and one $M$-cycle passing through the algebraic
variety, corresponding to $\mathcal{I}_{2}$. Therefore, before taking
the symmetry into account, there are 3 independent $M$-cycles, which
corresponds to 3 master integrals. The symmetry of the integral $z_{3}\leftrightarrow z_{6}$
results in $z^{\left(1\right)}\leftrightarrow z^{\left(2\right)}$
, therefore, the two contours passing through $z^{\left(1\right)}$
and $z^{\left(2\right)}$ are symmetry equivalent. Thus, the account
of symmetry relations leaves us with 2 master integrals.

\section{Conclusion}

We have shown that the number of master integrals with a given set of denominators can be determined by examining the critical set of the polynomial $G=U+F$, where $U$ and $F$ are two Symanzik polynomials entering the parametric representation. Alternatively, one can consider critical set of the polynomial $P\left(D_{1}=0,\ldots D_{M}=0,D_{M+1},\ldots D_{N}\right)$ entering Baikov representation. In the case of isolated proper critical points, the number of master integrals is just the number of proper critical points counted with multiplicity. This equality follows from the construction of the independent integration contours as Lefschetz thimbles attached to the critical points. It seems that this geometrical construction should have some other applications beyond a simple counting of the master integrals. We have presented a simple \emph{Mathematica} package \texttt{Mint} which automatically finds
the number of master integrals with a given set of denominators.

\subsection*{Acknowledgments} 

This work is supported in part by RFBR grants
11-02-00792, 11-02-01196, 12-02-33140, 13-02-01-023 and the grant
of Ministry of Education and Science. A.P. would like to thank Anatoly
Pomeransky for many useful discusions. We are grateful to Pavel Baikov for explaning some details of Ref. \cite{Baikov2006325} and to York Schr\"oder for a useful feedback on the \texttt{Mint} package.

\bibliographystyle{JHEP}
\providecommand{\href}[2]{#2}\begingroup\raggedright\endgroup

\end{document}